\documentclass[10pt,final,journal,twocolumn]{IEEEtran}
\usepackage[utf8]{inputenc}
\usepackage{cite}
\usepackage[cmex10]{amsmath}
\usepackage{amsmath}
\usepackage{amsfonts}
\usepackage{amsthm}
\usepackage{algorithm}
\usepackage{multirow}
\usepackage{multicol}
\usepackage{stfloats}
\usepackage{multicol,multienum}
\usepackage{multirow}
\usepackage[caption=false,font=footnotesize]{subfig}
\usepackage{wrapfig}
\usepackage{color}
\usepackage{graphicx}
\usepackage{xspace}
\usepackage{epstopdf}
\usepackage{suffix}
\usepackage{mathtools}

\begin{document}

\title{Federated Transformed Learning for a Circular, Secure, and Tiny AI
\thanks{$^{1}$Alan Turing Institute, UK; $^{2}$Cranfield University, UK; $^{3}$Bin Li is with Beijing University of Posts and Telecommunications, China; $^{4}$Birkbeck, University of London, UK.
$^*$Corresponding Author: wguo@turing.ac.uk.
The author wishes to acknowledge the Alan Turing Institute under the EPSRC grant EP/N510129/1, and Trustworthy Autonomous Systems (TAS) - Security Node under EPSRC grant EP/V026763/1. }
}
\author{
\IEEEauthorblockN{Weisi Guo$^{1,2}$, Schyler Sun$^{2}$, Bin Li$^{3}$, Sam Blakeman$^{1,4}$}} 

\maketitle

\begin{abstract}
Deep Learning (DL) is penetrating into a diverse range of mass mobility, smart living, and industrial applications, rapidly transforming the way we live and work. DL is at the heart of many AI implementations. A key set of challenges is to produce AI modules that are: (1) "circular" - can solve new tasks without forgetting how to solve previous ones, (2) "secure" - have immunity to adversarial data attacks, and (3) "tiny" - implementable in low power low cost embedded hardware. Clearly it is difficult to achieve all three aspects on a single horizontal layer of platforms, as the techniques require \textit{transformed deep representations} that incur different computation and communication requirements. 

Here we set out the vision to achieve transformed DL representations across a 5G and Beyond networked architecture. We first detail the cross-sectoral motivations for each challenge area, before demonstrating recent advances in DL research that can achieve circular, secure, and tiny AI (CST-AI). Recognising the conflicting demand of each transformed deep representation, we federate their deep learning transformations and functionalities across the network to achieve connected run-time capabilities.
\end{abstract}

\begin{IEEEkeywords}
machine learning; deep learning; adversarial machine learning; compression; XAI; 5G; 6G;
\end{IEEEkeywords}

\IEEEpeerreviewmaketitle

\section{Introduction}

As 5G networks roll out across the world, researchers are identifying the societal challenges for the next few decades in order to design the 6G networks that will serve as the digital backbone. Over 50 billion autonomous and IoT devices is set to be networked by end of the century, adding a hierarchy of distributed intelligence ranging from onboard embedded algorithms (e.g. dedicated purpose, sensitive to observations) to cloud intelligence orchestrating a range of services (e.g. episodic memory). As such, future 5G and 6G networks are likely to be increasingly integrated with the AI modules in a hyper-dense mass autonomous digital economy \cite{Li20}, where intelligent agents (from autonomous vehicles to data analytic engines) are complementing and supplementing human labour across a diverse range of local industrial, commercial, agricultural, and mobility services. In particular, many autonomous and vehicular systems have stringent AI service performance demands that range form ultra low-latency decision-making to complex bio-inspired memory architectures to avoid catastrophic forgetting. These cannot co-exist on common hardware platforms and requires a federated approach. 

\subsection{Challenges \& Federated Learning}
There are currently 3 key barriers to scalable mass autonomy across a wide range of applications: (1) AI catastrophically forgetting previous knowledge and re-learning at high computation and time costs \cite{background}, (2) AI vulnerability to adversarial data attacks \cite{AdvML}, and (3) AI cannot be implemented on front-end machines due to limited or unsustainable processing power scaling laws \cite{new1, Du19, Energy}. 

\begin{figure*}[t]
     \centering
     \includegraphics[width=0.9\linewidth]{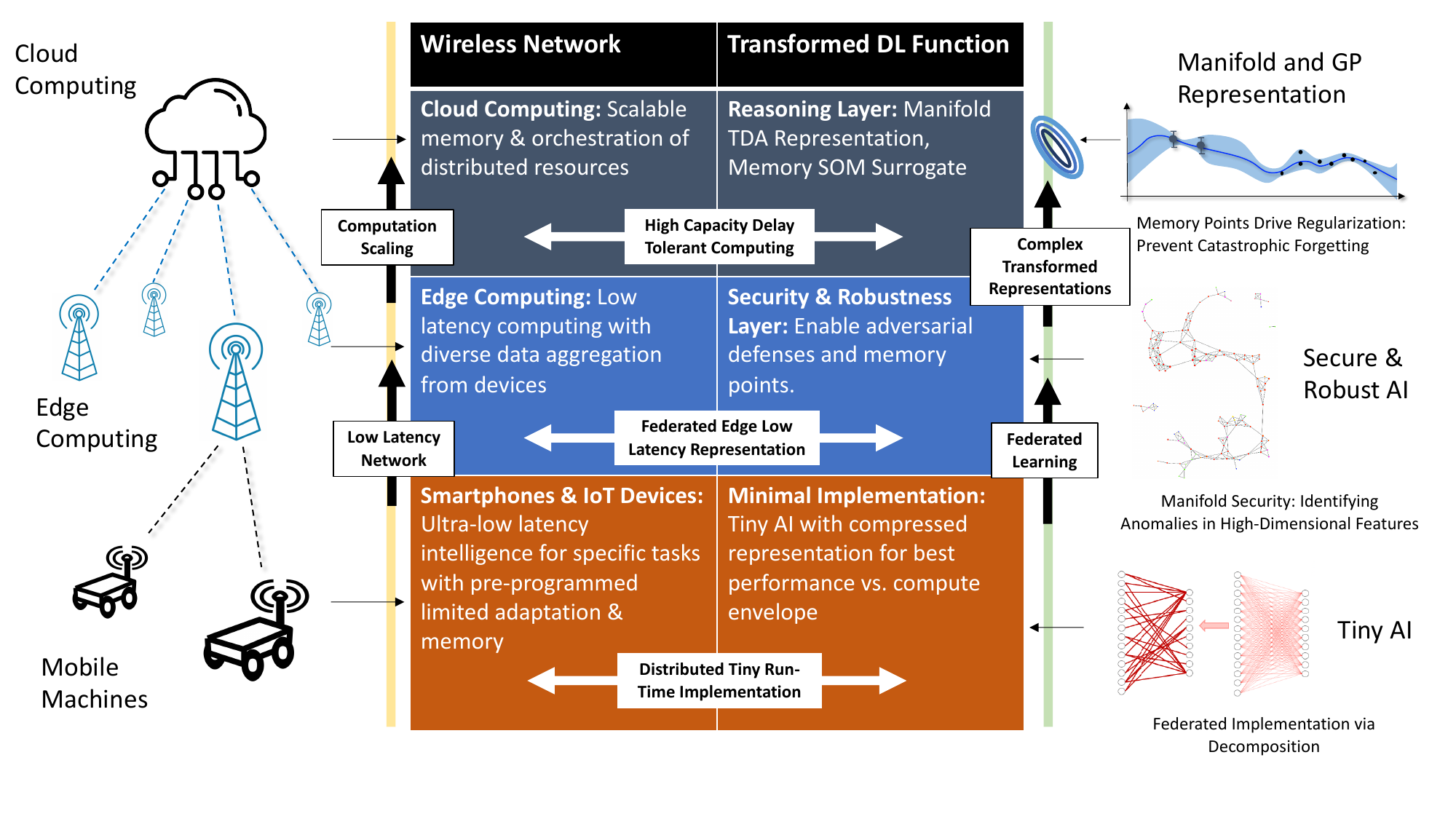}
     \caption{Digital ecosystem for Circular, Secure, and Tiny AI (CST-AI) with distributed functionalities across onboard-edge-cloud, and connected via wireless network: cloud based digital twins enable transformed representation of DL - which enables circular and robust AI; edge processing secure AI against adversarial attacks - implementing many of the transformed functionalities from cloud and aggregating data from diverse devices/machines; and onboard intelligence need real-time tiny AI for scalable capability.}
     \label{fig1}
\end{figure*}

It is widely recognised that these need to be addressed, but doing so on a single on-board or edge device incurs a high level of sophistication, compute energy requirement, and real-time data demand. In order to overcome these challenges, the community have made advances in the design of new AI algorithms and re-distributing its functionalities across the wireless network to leverage on various processing-communication capability trade-offs. One key advance is federated learning, where fragmented representations of the artificial neural network (ANN) is distributed across embedded and edge devices to enable scalable and decentralised or hybrid learning \cite{Fed20}. Whilst this addresses some of the above challenges in security and scalability, federated learning fundamentally computes in the classic ANN domain. That means many of the challenges related to security and catastrophic forgetting cannot be solved. Yet, if we are to transform the ANN into more complex representations and develop computational twins for various purposes, then this requires re-distributing those \textit{transformed DL models} across the network for efficient computation. The diverse AI requirements differ significantly for the different ANN representations that serve different AI purposes via a wireless network - see Figure \ref{fig1}. 

\subsection{Novelty \& Contribution: Federating Transformed Deep Representations}
The opportunity here, is to develop wireless network enabled AI ecosystems that achieve CST-AI through federating the transformed representations. We will show how a wireless network with agile slicing is necessary: 
\begin{enumerate}
    \item Section II: A networked ecosystem to achieve CST-AI via federating transformed DL functionalities across the wireless network with agile slicing. We review the key concepts and why a wireless network must federate these functionalities, before explaining the detailed methods below.
    \item Section III: Circular DL aims at eliminating catastrophic forgetting through using parallel memory agents (unsupervised) or memory points (supervised) to enforce learning - significantly reducing computational costs and improving performance;
    \item Section IV: Secure DL through adversarial learning, statistical certificates, and manifold defence - offering either practical and theoretical benefits;
    \item Section V: Computational scaling of deep learning implementation using novel compression and surrogate model techniques to overcome hardware scaling issues and enable practical embedded intelligence;
    \item Section VI: Summary and open challenges;
\end{enumerate}
We hope this review enables readers to think about the joint challenges between AI for mass autonomy and wireless network design. 

\begin{table*}[!t]
	\centering  
	\caption{Federated Transformed Learning Functionalities distributed across onboard, edge, and cloud capabilities on a wireless network.}
	\label{table2}
	\begin{tabular}{|c|c|c|c|}  
		\hline 
		\textbf{Deep Transforms}&\textbf{Circular}&\textbf{Secure}&\textbf{Tiny}\\
		\hline
        \textbf{Cloud-side} 
        &Memory Surrogates \cite{Blakeman}, Memory Points \cite{CL2}    
        &Statistical \cite{AdvML2}, Manifold Transform \cite{AdvML6}
        &On-the-fly Decomposition \\
        \hline
        \textbf{Edge-side} 
        &Temporal Difference \cite{Blakeman}, GP Kernel Update \cite{CL2}    
        &Anomaly Detection \cite{AdvML2, AdvML6}
        &Federated Learning \cite{Fed20} \\
        \hline
        \textbf{Machine-side} 
        &DL Model Update   
        &N/A
        &Post-Training Decomposition \cite{Du19}  \\
        \hline
	\end{tabular}
\end{table*}

\section{Wireless Network Ecosystem}
\label{section2}

We envisage that future AI modules that serve mass-automation (vehicles to factory) will distribute different services across the wireless network ecosystem in order to meet stringent AI-based Quality-of-Service (QoS) metrics: low AI model update latency, computational scalability \cite{Du19}, adversarial AI security, AI explainability \cite{Guo20}, and circular AI sustainability. 

\subsection{On-board AI: Tiny}
Starting at the bottom device level, there is urgent need to make DL algorithms achieve run-time performance on embedded platforms. Recent research \cite{new1} indicate a roughly >20$\times$ gap in storage size, and >10$\times$ gap in power consumption, between the DNN demands and the achievable computational capability on-board or at the edge. For example, a micro-controller DSP/FPGA sub-system typically has a power budget of 10mW and a storage size of 100kB on-chip memory, whist the corresponding DNN require 2MB of memory in weights and a significantly higher run-time energy demand. Whilst federated learning or cloud access critical, but both will incur latency and wireless access challenges. Therefore, the lightweight energy efficient DL paradigm \cite{Du19} becomes critical in attempting to meet the hardware gap that exists for the edge learning explained in detail in Section\ref{section5}.  

\subsection{Edge AI: Secure \& Robust}
Edge AI coordinates different aspects of the connected vehicles through data aggregation for federated learning \cite{Fed20}. With over 50 billion devices set to be networked by end of the century, there is an unprecedented opportunity for malicious attacks. A particular danger is a false identity attack which injects poisonous data into the edge AI. The urgent need is to make the edge AI secure against spoofing and adversarial data attacks \cite{AdvML} (see Section\ref{section4}). Other edge roles include aggregating not only training data in federated learning, but also the transformed representation data from cloud to inform security and robustness protocols. This leads us to the final upper layer of the ecosystem, where the transformed representations take place.

\subsection{Cloud AI: Transformed Representation for Circular Economy}
The transformed representations on cloud enable a circular digital economy, because by understanding DL's reasoning in a transformed domain, we can dramatically reduce the likelihood of re-training/updating or re-purposing DL algorithms. As far as we are aware, this is the first time "circular AI" has been proposed, e.g. re-purposing the AI efficiently across a wide range of tasks and scenarios. Taking an example from NLP task training, the carbon footprint cost (CO$_2$ equivalent) of tuning a parsing pipeline is 35T and a large transformer is 284T \cite{Energy}. This is roughly equivalent to the equivalent carbon footprint generated by 40 and 313 passengers on a medium haul flight respectively, or 0.6 and 5 petrol cars in their typical life time. The research frontier (see Section\ref{section3}) addresses this emerging energy efficient AI scalability challenge by creating methods that inject "memory" into the learning agent or identify deep feature space patterns that can be exploited for greater robustness. The application example is that this cloud-based digital twin responds to the diverse range of observations seen by robots and makes ecosystem level decisions in resource allocation, task assignment, and risk taking. These often require memory of previous lessons learnt to avoid catastrophic forgetting, and repeating lessons in the field. 

\subsection{Agile Network Slicing to Achieve CST-AI}
As shown in Figure \ref{fig1}, this CST-AI ecosystem creates stringent demands on the wireless network because it requires different AI transformed functionalities to be distributed, and to have their data streams and QoS demands streamed across the network. Network slicing is an integral part of the current 3GPP R16 and largely depends on explicit and timely knowledge on the demand and network environment condition. This enables a range of optimisation solutions ranging from classic optimisation to DL. However, many of the AI modules on-board and at edge are not known and change dynamically. The Radio Resource Management (RRM) need to allocate resources, reacting on the millisecond order in mass autonomy cases \cite{Li20}. This requires that a slicing is conducted in the absence of system state information while performance safeguards are kept across the run time trajectory. This leads to the need to create run-time slicing for AI demands. As shown in \cite{slice}, directly solving an off-line slicing solution is not feasible and a DL approach with stability guarantees can enable the system to learn a safe slicing solution from both historical records and run-time observations. To facilitate the imagination of the reader, a range of transformed DL functionalities and their QoS requirements are given in Table \ref{table2}. 

Having established the agile network slicing ecosystem to enable CST-AI, we now detail the specific innovations in different areas of the transformed representations and where they sit in the networked ecosystem.

\section{Circular AI}
\label{section3}

Overcoming catastrophic forgetting is the key to the continual learning in our proposed circular AI approach. This requires DNNs to not only hold its performance on previous datasets/tasks but also be capable of adapting to new datasets/tasks without significant repetitive training \cite{background}. 

\begin{figure}[t]
     \centering
     \includegraphics[width=0.95\linewidth]{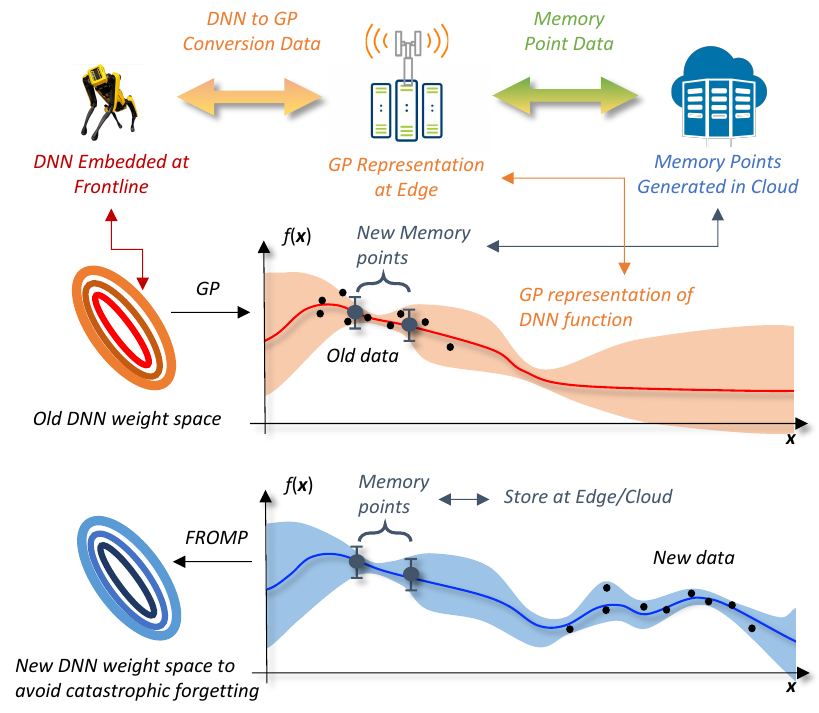}
     \caption{Circular Supervised AI using function regularization via memory points: embedded front-end DNNs are represented by GPs at the edge with cloud computing of complex memory points.}
     \label{fig2}
\end{figure}

\subsection{Supervised Learning with Memory Points}
Regularization over the DNN \textit{weight space} or \textit{function space} are state-of-the-art approaches in avoiding catastrophic forgetting. Weight regularization \cite{background} fine-tunes the DNN parameters in further datasets/tasks training with parameter-regularisation so that the information learnt from previous datasets/tasks can be retained to some extent. However, due to the complex non-linear relationship between NN parameters and outputs, such direct approaches cannot theoretically guarantee its robustness to new arbitrary datasets.

\textbf{Transformed Representation as GPs:} More recently, functional-regularisation (FR) directly regularises the equivalent function space of the DNN, by modeling the DNN as a Gaussian Processes (GP) - owing to the theoretical connections relating DNNs and GPs \cite{DNN2GP}. As shown in Figure \ref{fig2}, the GP representation of the DNN weight space is the functional prior for the next new datasets/tasks training, which becomes approximately locally tunable. Hence, it is able to fit the new datasets/tasks with regularisation on its outputs based on previous memory points \cite{CL2}. Leveraging on the reverse mapping from GPs to DNN parameters, methods such as FROMP \cite{CL2} can translate functional prior regularisation into new DNN weights. An inevitable issue for state-of-the-arts methodologies is the infeasible computational complexity, whereupon, a number of approximations and assumptions are addressed in current applications. Therefore it becomes critical to develop a distributed computing system across the wireless network as shown in Figure \ref{fig2}. The front-end supervised learning is likely to be onboard machines/robots, whilst the complex GP representation and update process requires edge or cloud computational power. This highlights how the wireless network is essential in delivering circular AI.

\subsection{Reinforcement Learning with Surrogate Mind Map}

Deep Reinforcement Learning (DRL) has made it possible for AI systems to continuously interact with their environment. Many of the advances in DRL have been inspired by the human brain and its ability to transform high-dimensional sensory input into action. Importantly, the human brain appears highly adept at overcoming catastrophic forgetting as it continually learns new tasks without forgetting what it has previously learnt. It has been proposed that one of the reasons the brain is able to overcome catastrophic forgetting is due to its use of complementary learning systems. More specifically, it has been suggested that the brain uses episodic memory to rapidly store individual experiences and semantic memory to learn generalizations across many experiences. Crucially, this sampling procedure is done in an interleaved fashion so that new information is combined with old information and catastrophic forgetting is prevented.

\textbf{Transformed Representation as Mind Map:} Seminal work in DRL, such as the Deep Q-Network, has taken inspiration from this brain architecture to successfully train DNNs to perform RL. In the case of DRL, the episodic memory system typically corresponds to a table of past experiences while the semantic memory system corresponds to a DNN. Experiences are constantly added to the table and the DNN is trained intermittently by sampling at random from the table. With respect to our recently proposed Continuous Temporal Difference Learning (CTDL) \cite{Blakeman}, the DNN can be used for function approximation at the edge while the table of past experiences can be stored in the cloud where more memory resources are available. If a copy of the DNN is also stored in the cloud then parameter updates can be calculated server-side and then passed to the edge using a wireless connection.

\begin{figure}[t]
     \centering
     \includegraphics[width=0.95\linewidth]{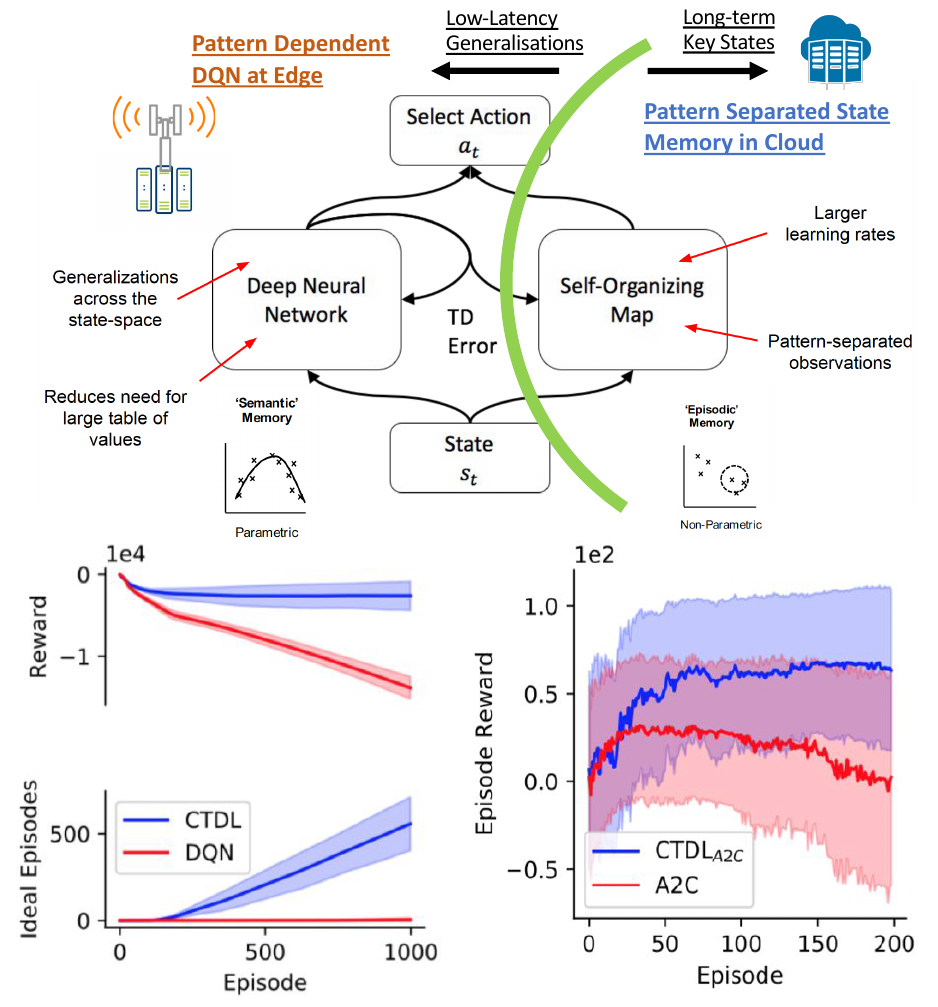}
     \caption{Circular Reinforcement AI using CTDL: (top) edge-based pattern dependent DQN is wireless connected to a long-term pattern independent SOM memory, reducing catastrophic forgetting through temporal differences (TD), (bottom) CTDL achieves superior performance in a wide range of RL tasks against standard DQN and Advantage Actor Critic (A2C) algorithms.}
     \label{fig3}
\end{figure}

CTDL utilizes the errors produced by the DNN to strategically decide which experiences to store in episodic memory, which in this case is represented as a Self-Organizing Map (SOM) (see Figure\ref{fig3}). This reduces the memory resources needed for the episodic memory system because the SOM only stores experiences in long-term memory that the DNN is poor at evaluating.  The experiences in long-term memory therefore reflect problematic data points that are not amenable to function approximation. CTDL achieves superior performance in a wide range of RL tasks against standard DQN and Advantage Actor Critic (A2C) algorithms. An additional benefit of CTDL is that the experiences stored in long-term memory can be used for prediction using non-parametric approaches. These non-parametric predictions are immune to catastrophic forgetting for as long as the experiences are held in long-term memory. From the perspective of CST-AI, the edge can use the DNN for parametric function approximation, while the cloud can be used for non-parametric predictions when regions of the environment are visited that the DNN is poor at evaluating.

\section{Security \& Trust}
\label{section4}

Mass autonomy in cross-sectoral applications (e.g. mobility as a service, multi-infrastructure smart grid) often require consumer and industrial data sets that are heterogeneous and high-dimensional. This causes uncontrolled systematic noise resulting from high dimensional noise or adversarial data attacks - both of which are difficult to expose at the high dimensional levels of the DNN \cite{AdvML}. The research is divided into developing both real-time data-driven defences, and statistically grounded certificate defences. 

\subsection{Adversarial Training}
The most practical method set is data-driven defences that provide a wide of training to the DNN to add a degree of adversarial immunity. In adversarial training, robust stochastic gradient descent (SGD) \cite{AdvML2} is one approach that tackles corrupted data or gradients during the training phase. Robust SGD with adversarial training at gradient $X$ is conducted by checking for adversarial examples $X^{*}$. However, this does not effectively deal with real time backdoor access to training data that add both data artefacts and mislabels. Whilst this empirical approach do not offer guarantees or certificates, they present implementable and real-time solutions to real problems. 

\begin{figure}[t]
     \centering
     \includegraphics[width=0.95\linewidth]{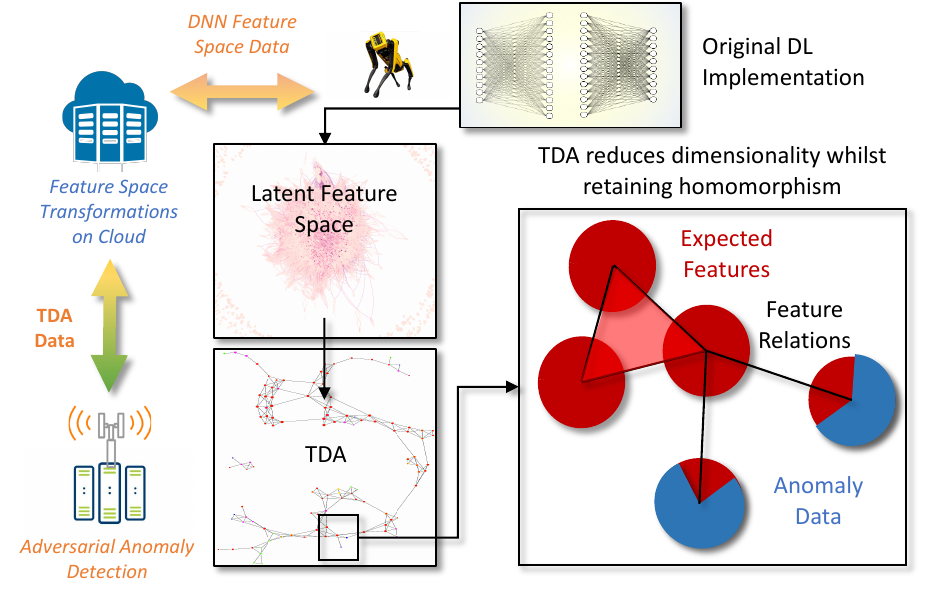}
     \caption{Manifold defence using TDA to identify anomaly data features latent in the high-dimensional space with nonlinear inter-feature relations.}
     \label{fig5}
\end{figure}

\subsection{Certificate Filters}
Certified defences offer proofs to what attacks can be countered using statistical guarantees integrated into the DNN. Traditionally, in low dimensional data, we can identify corruption/noise through covariance checks via highest eigenvalue and remove the projection for real time cleaning. Filters can be implemented in the final representation layer of the DNN. This provides a statistical certificate against adversarial noise injections. This becomes more challenging at higher dimensions, especially with mixed data types and mixed adversarial statistics (e.g. higher moments). Other certified defences that might not operate in real time include: (1) randomised smoothing with soft classifiers \cite{AdvML5}, and (2) manifold based defences to identify data topology anomalies \cite{AdvML6}. 

\textbf{Transformed Representation as Feature Topology:} The latter method of manifold defences has gained significant attraction lately. By comparing the feature distribution difference between trusted data and full data in the DNN, one requires dimension reduction. Topological data analysis (TDA) can identify high-dimensional anomalies via persistent homology - see Figure \ref{fig5}. Through identifying simplicial complexes and multidimensional persistence, TDA can give a more comprehensive structure of the data feature distribution with intrinsic clusters by preserving local relationships of the high-dimensional feature space rather than conventional Principal component analysis (PCA) and Multidimensional scaling (MDS) methods - which do not capture any preserved data structure. The complexity of TDA implementation means it may very well be implemented on the cloud, but have the anomalies used to achieve security at the edge.

\section{Tiny AI}
\label{section5}

As discussed previously in circular AI, the on-board training of DNNs at the edge is critical for low latency multi-task applications such as autonomous piloting, mission critical diagnostics, and agile manufacturing. In many such applications, the on-board or edge hardware has limited storage size, computational capability, energy budget, and communication capacity. It has to transition between different tasks, learning or updating its DNN on the fly or in run time. 

\subsection{Lightweight DNN Methods}

\begin{figure}[t]
     \centering
     \includegraphics[width=1\linewidth]{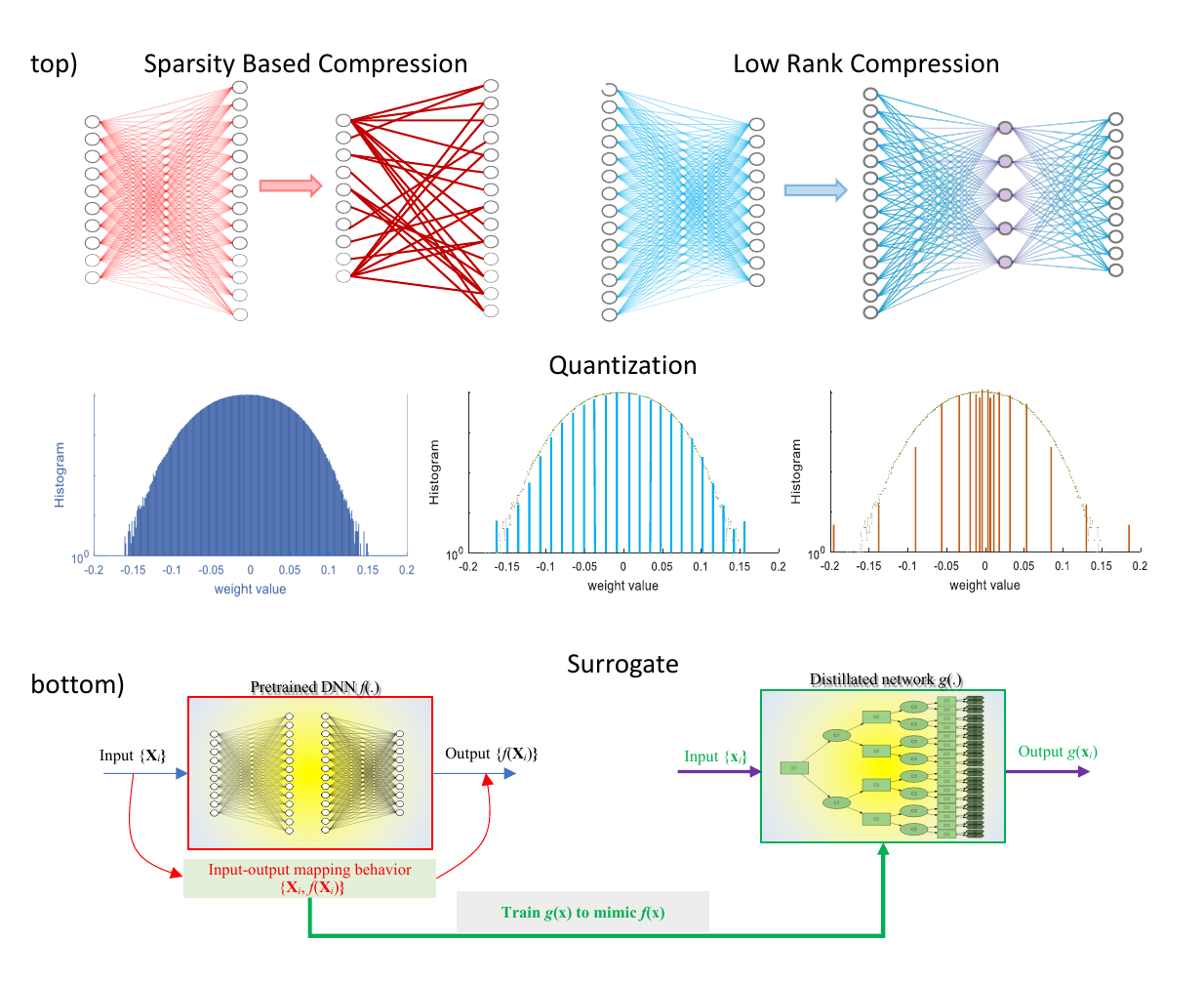}
     \caption{Tiny AI: Compression, Quantization, and Distillation network methods.}
     \label{fig6}
\end{figure}

The state-of-the-art methods combine joint hardware design (e.g. near-data processing, non-von Neumann architectures, systolic array architecture) and DNN compression. Parallel to the hardware innovations, the advances in lightweight algorithms have boosted the widespread use of DNN, which are designed to abstract a compressed DNN representation via \cite{Du19}: (1) network pruning/compression, (2) weight quantization, and (3) network distillation, see Figure \ref{fig6}. 

\textbf{Transformed Representation via Compression:} The first method aims to compress a pre-trained network, by exploiting two dominant features of the learned weights, i.e. the sparsity and the low-rankness. The former reserves the network connections with large weights (see Figure \ref{fig6}top-left), without significantly degrading the prediction/classification accuracy; whilst the latter attempts to decompose a large pre-trained weight matrix into multiple small matrices (see Figure \ref{fig6}top-right), thereby reducing the storage and computation burden. According to some previous works, the pre-trained network may be reduced by 2-10 folds via such post-training compression methods. 

The second method focuses on the machine precision adaptation of network weights (see Figure \ref{fig6}middle). By replacing the float-point weights (e.g. in 32 bits) with the low-resolution fixed-point weights (e.g. in 8 bits), the storage size can be effectively reduced. Similar to the aforementioned weight compression method, there also exists a compromise between the network accuracy and the weight reductions. In practice, one may assume the test accuracy would be slightly degraded by <1\%, and then exploits various quantization schemes, e.g. uniform, logarithm or other user-defined schemes. 

\textbf{Transformed Representation via Distillation:} The distillation network method tends to identify one smaller surrogate network to mimic the input-output mapping behaviors learned by an originally pre-trained DNN; such methods may include the local approximation for a small subset of input data, and the model translation for the entire dataset (see Figure \ref{fig6}bottom). A translated network may be implemented by decision tree, Bayesian graphs or even regressions. As demonstrated recently, such model translation techniques would achieve a largely reduced network, which is thus much easier to deploy in the edge scenarios and, more importantly, becomes even easily explainable \cite{Guo20}.  
            
\section{Conclusions and Open Challenges}
\label{section6}

We know AI is going to change the way autonomous systems and wireless networks interface. What we don't yet know is exactly what the frontier research challenges are in this space. Here, we propose that "circular", "secure", and "tiny" are the 3 critical aspects of this challenge space. We detailed the cross-sectoral motivations for each area, before demonstrating recent advances in AI research that can achieve circular, secure, and tiny AI (CST-AI). Recognising the conflicting demand of each attribute, we distribute their functionalities across the network to achieve run-time capabilities using agile network slicing to achieve a future fit for mass autonomy.

That being said, there is a long way to go before we connect connect diverse AI requirements and implementations seamlessly into a wireless network. We believe the key open challenges are:
\begin{enumerate}
    \item Compress-on-the-Fly: building tiny AI modules that compress their architecture dynamically on the fly whilst training and operating, will be significantly better than post-training compression methods. This currently does not exist and requires new ways of integrating back-propagation with compression.
    \item Joint Communications and AI Defence: building secure edge intelligence that is jointly secure from a communication and AI perspective. Drawing inspiration from physical layer and public key security, we can develop methods to authenticate meaningful data via TDA manifold methods.
    \item Circular AI with Human Knowledge Integration and Explainability: building circular AI that can speak and interact with human experts so that catastrophic forgetting mitigation extends to a wider range of tasks and a deeper mutual understanding of the problem domain. This is done by translating human knowledge into memory points to regularize DNN functions, whilst explaining back the DNN's reasoning to the human user via semantic or algebraic methods \cite{Guo20}.
\end{enumerate}
Achieving these objectives will enable the human society and machine world to be more integrated, facilitating cross learning and a smarter living.

\bibliographystyle{IEEEtran}
\bibliography{Ref} 

\begin{IEEEbiography}{Weisi Guo} (S07, M11, SM17) received his MEng, MA, and Ph.D. degrees from the University of Cambridge, UK. He is Chair Professor of Human Machine Intelligence at Cranfield University. He has published over $170$ papers and is PI on over $\pounds4$m of research grants. His research has won him several international awards (IET Innovation 15, Bell Labs Prize Finalist 14 and Semi-Finalist 16 and 19). He was a Turing Fellow at the Alan Turing Institute and is a Fellow of Royal Statistical Society.
\end{IEEEbiography}

\begin{IEEEbiography}{Schyler Sun} (S07, M11, SM17) is a PhD student at Cranfield University, and works on mathematical proofs for explainable deep learning.
\end{IEEEbiography}

\begin{IEEEbiography}{Bin Li} (S07, M11, SM17) is an associate professor at BUPT and works in designing front end tiny AI implementations using novel signal processing decompositions.
\end{IEEEbiography}

\begin{IEEEbiography}{Sam Blakeman} is a PhD student at Birbeck College London, and has worked on a wide range of human brain inspired reinforcement learning agent design.
\end{IEEEbiography}

\end{document}